# Microstructural characteristics, atomic-scale features, and growth mechanisms of deuterides (hydrides) in hafnium

Di Wang, Catriona M. McGilvery, James O. Douglas, Siyang Wang*

Department of Materials, Royal School of Mines, Imperial College London, London, SW7 2AZ, UK

*Corresponding author: siyang.wang15@imperial.ac.uk

## Abstract

Hafnium hydride is a promising material for next-generation nuclear reactors, particularly as control rods for fast fission and shielding in fusion systems. The material's intrinsic brittleness encourages its use in the form of hydride-metal composites, where the functional and mechanical performance is strongly influenced by the multiscale structure of hydride-matrix interfaces. In this study, we employ a suite of microscopy techniques, including scanning electron microscopy with electron backscatter diffraction, transmission electron microscopy with electron energy-loss spectroscopy, and atom probe tomography, to investigate the deuteride-matrix interfaces in a deuterium-charged Hf alloy. We characterise their structure and chemistry, extracting key information including the deuteride-matrix crystallographic orientation relationship, microstructural features, misfit-induced dislocation distributions, electron energy-loss characteristics, and the segregation of oxygen during deuteride growth. These findings help clarify the mechanisms of interface evolution and may contribute to improved understanding of hydride-metal systems, with potential relevance for their processing, performance, and design in nuclear applications.

## 1. Introduction

Hydrogen and hafnium are two elements closely related to the nuclear industry. Hydrogen is effective in slowing down fast neutrons - high-energy neutrons produced during fission - through elastic scattering, converting them into thermal neutrons [1]. In the vast majority of fission reactors, water serves as a neutron moderator (alongside its role as a coolant) due to the moderation capability of hydrogen. Meanwhile, hafnium exhibits excellent neutron absorption properties, particularly for thermal neutrons, making it suitable for control rods that regulate or stop chain reactions when necessary [2–8]. As a result, hafnium-based control rods are especially utilised in naval reactors [2], where precise and reliable neutron regulation is essential. Beyond nuclear applications, hafnium also finds potential use in optical hydrogen sensors [9], as $HfN_{0.4}$ in ferroelectric memories [10], and as an alloying element in materials for hydrogen storage [11,12].

There has been growing research interest in hafnium hydride, leveraging the combined properties of hydrogen and hafnium, particularly for applications in advanced nuclear reactor types. For example:

- Fast fission reactors: In these reactors, while fast neutrons are used to sustain the fission process, control rods can benefit from hafnium hydride due to its dual properties. Hydrogen moderates the neutrons, slowing them down to thermal energies, where they can be more effectively absorbed by hafnium. This combination enhances the efficiency of neutron absorption for reactivity control, which has driven ongoing research into hafnium hydride as a control rod material [8,13–25].

- Fusion reactors: Hafnium hydride has gained interest in recent years as a promising candidate for shielding materials in fusion reactors [26,27]. Fusion reactions, particularly deuterium-tritium fusion, produce both high-energy neutrons and gamma radiation, which can damage reactor components and reduce operational efficiency. The combination of hafnium's gamma ray attenuation and hydrogen's neutron moderation helps protect critical systems and extend the lifetime of reactor materials.

The Hf-H(D) binary system is generally analogous to those of other Group IV elements such as Zr and Ti [28]. Three stable phases at atmospheric pressure have been observed: the hexagonal close-packed (HCP) α-Hf phase, the face-centred cubic (FCC) δ-Hf(H/D)$_x$ phase ($x \gtrsim 1.5$), and the face-centred tetragonal (FCT) ε-Hf(H/D)$_x$ phase ($c/a < 1$, as $x$ approaches 2) [29–34]. Two additional phases have been reported: the FCT γ-Hf(H/D)$_x$ phase ($c/a > 1$, $x \approx 1$), which is less stable than the δ phase [23], and the FCC δ'-Hf(H/D)$_x$ phase, characterised by a more ordered arrangement of H/D atoms compared to the δ phase [23,25]. Furthermore, it has been observed that as the H concentration in the ε phase increases, the lattice parameters evolve ($a$ increases while $c$ decreases), resulting in greater structural anisotropy [35]. Beyond phase compositions, similarities and differences between Zr and Ti hydrides in their interface structures and hydride behaviour have been discussed in prior literature [36]. Such systematic comparison with the Hf-H system, however, is missing to the best of our knowledge, presumably due to the limited availability of detailed characterisation studies of Hf-HfH$_x$ interfaces.

The intrinsic brittleness of bulk hafnium hydride [29,30,37] may pose a significant challenge to maintaining its structural integrity in harsh reactor environments over long term. One approach currently being explored to address this issue is the development of composite materials in which hydrides are embedded within an iron/steel matrix [38,39]. The metallic matrix would provide ductility and mechanical support, counteracting the brittleness of the hydride. This approach could result in materials better able to withstand the service conditions compared to pure hydride.

Composite materials in which hafnium hydrides are entrained within a hafnium metal matrix, however, have received little attention to date. At the industrial scale, potential manufacturing routes could include powder metallurgy [39,40] and gaseous hydrogen annealing of hafnium metal [41]. For example, Ito *et al.*, [18] showed that prolonged annealing of metallic hafnium (99.9% purity) in a hydrogen atmosphere produced crack-free single-δ-phase bulk hafnium hydride. This suggests that careful adjustment of annealing parameters (e.g., shorter duration) may enable fabrication of Hf-HfH$_x$ composites *via* partial hydrogenation. The performance of such composites is likely governed by the behaviour of Hf-HfH$_x$ interfaces, yet their fundamental characteristics remain largely unexplored. Indeed, hafnium is generally believed to form hydrides similar to those in zirconium [37], presumably owing to similarities in their chemical nature, phase diagrams with hydrogen, and mesoscale hydride structures. However, whether the structural, crystallographic, and chemical features of Hf-HfH$_x$ interfaces mirror those of Zr-Zr$_x$ interfaces remains an open question.

In this work, therefore, we focus on the characterisation of Hf-HfH$_x$ interfaces using advanced microscopy techniques. The material examined was produced through a laboratory-scale hydrogen charging method previously applied to Zr [42]. Although this differs from the potential industrial approaches mentioned above, the characteristics of the resulting Hf-HfH$_x$ interfaces are expected to be comparable, especially to those formed during gaseous hydrogen charging, where hydrides likewise precipitate during cooling from elevated temperatures.

Using electron backscatter diffraction (EBSD), transmission and scanning transmission electron microscopy (TEM/STEM), electron energy-loss spectroscopy (EELS), and atom probe tomography (APT), we reveal details such as grain orientation, chemical distribution, and atomic-scale structure. These findings provide new insights into the mechanisms governing hydride growth, and form a foundation for understanding how these materials might behave under the demanding conditions encountered in advanced reactors.

## 2. Experimental methods

### 2.1. Sample preparation

A commercially available Hf alloy, provided in the form of a drawn rod by Goodfellow Cambridge Ltd., was used in this study. The supplier's chemical composition data for the alloy are listed in Table S1. A sample was sectioned from the rod and annealed at 1000 °C for 240 h to promote recrystallisation and grain growth, resulting in the formation of large equiaxed grains.

Electrochemical deuterium charging of the sample was carried out using the same method previously employed for H/D charging in Zr [42–46]. We chose to use deuterium charging to minimise overlap with background hydrogen peaks and to distinguish introduced deuterides from any focused ion beam (FIB)-induced hydride artefacts in subsequent APT analysis [46,47]. The process involved electrochemically charging the sample in a solution of 1.5 wt.% $D_2SO_4$ in $D_2O$, applying a current density of 2 $kA/m^2$ at 65 °C for 24 h. This led to the formation of deuteride layers on the sample surfaces. The sample was then annealed at 400 °C for 5 h to promote the diffusion of deuterium from the surface into the bulk. Finally, slow furnace cooling at 1 °C/min induced the precipitation of deuterides throughout the microstructure.

Following deuterium charging and homogenisation, the sample was ground using SiC abrasive papers (up to P4000 grit, ~5 µm grain size), mechanically polished with colloidal silica (0.25 µm grain size), and finally electropolished in a solution of 10 vol.% perchloric acid in methanol at -50 °C for 90 s with an applied voltage of 30 V. This procedure was used to remove the surface stress layer and enhance the quality of the subsequent EBSD characterisation.

### 2.2. EBSD

EBSD characterisation was conducted using a Thermo Fisher Scientific (TFS) Quanta 650 scanning electron microscope (SEM) equipped with a Bruker eFlashHR (v2) EBSD camera. The beam acceleration voltage was set to 20 kV with a probe current of ~10 nA. EBSD patterns were binned from a native resolution of 1600 × 1200 pixels to 320 × 240 pixels, with an exposure time of 21 ms. Large maps were acquired to characterise the microstructure of the large-grain hafnium alloy prior to D charging, while smaller maps focused on individual deuteride packets, using step sizes of 5 µm and 200 nm respectively.

### 2.3. TEM/STEM and EELS

Thin foil specimens for TEM/STEM experiments were prepared from the growth front of a grain boundary deuteride using *in situ* Xe FIB lift-out on a TFS Helios 5 Hydra DualBeam microscope. Xe FIB, rather than Ga FIB, was used as it was found to reduce the formation of FIB-induced hydrides in the sample.

TEM/STEM imaging and EELS analyses were carried out on a TFS Spectra 300 STEM equipped with a Cs probe corrector and monochromator, using a beam acceleration voltage of 300 kV. EELS data were acquired on a Gatan Continuum K3 HR spectrometer on the Continuum complementary metal-oxide semiconductor (CMOS) camera. A 1 mm Gatan Imaging Filter (GIF) entrance aperture was used giving a convergence angle of 18.3 mrad and collection angle of 24.5 mrad to the spectrometer. The beam current through a hole was 150-160 pA as measured on the fluorescent screen. Under acquisition conditions the full-width at half-maximum (FWHM) of the zero-loss peak (ZLP) was 0.8-0.9 eV. Spectrum images (SI) were acquired in dual EELS mode with an offset of 15 eV between low-loss (LL) and high-loss (HL) data sets to improve signal to noise in the plasmon region of interest. LL and HL data were acquired at 1 ms/px and 10 ms/px respectively with 8x8 sub-pixel scanning enabled. Pixel spacing was 2-5 Å and an energy dispersion of 50 meV/ch was used.

The data sets were aligned to the ZLP. The deuteride and matrix regions were identified from the energy of the plasmon peak and maps were generated using multiple linear least squares fitting (MLLS) using 'standard' spectra from each region. Spectra showing the plasmon region were taken from the same number of pixels and have not had any further data processing carried out on them.

## 2.4. APT

APT specimens containing deuteride-matrix interfaces were fabricated site-specifically using *in situ* Xe plasma FIB lift-out on the Hydra microscope, following the procedure detailed by Thompson *et al.* [48]. APT measurements were performed using a Cameca LEAP 5000 XR atom probe microscope in laser pulsing mode, with a specimen base temperature of 60 K, a laser pulse rate of 60–150 kHz, laser energy of 60 pJ, and a detection rate of 0.1–0.3%. Data analysis was conducted using Cameca's integrated visualisation and analysis software (IVAS) in the AP Suite 6.3 toolkit.

## 3. Results

### 3.1. General microstructure of the materials

Figure 1(a) shows the microstructure of the Hf alloy after heat treatment at 1000 °C for 240 h. The material consists of equiaxed grains, and the texture resulting from drawing can be observed in the pole figures in Figure S1. After D charging and homogenisation, deuteride packets formed at grain boundaries (Figure 1(b)). These intergranular deuterides were subsequently characterised in detail with electron microscopy and atom probe tomography.

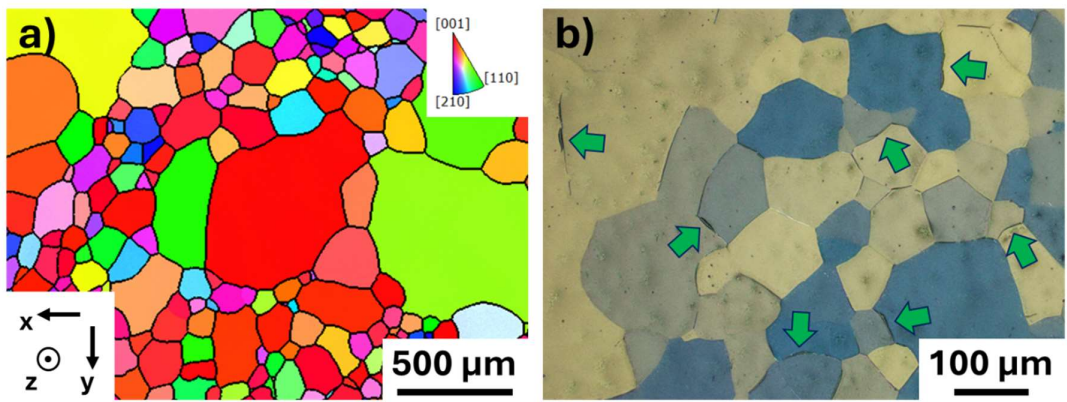

Figure 1 (a) Inverse pole figure (IPF)-*z* map of the hafnium alloy after heat treatment at 1000 °C for 240 h. (b) Polarised light optical micrograph of the material after deuterium charging and homogenisation, where deuteride packets can be observed at grain boundaries. Some intergranular deuerides are marked with the green arrows.

## 3.2. EBSD characterisation of intergranular deuteride packets

The EBSD results of an intergranular deuteride packet are shown in Figure 2, with the deuteride phase indexed as δ-deuteride. The crystal orientation map (Figure 2(a)) shows that the turquoise-coloured deuteride packet is predominantly located along the grain boundary. However, in two distinct regions (the two protrusions in the deuteride on the top left and bottom right of the map), it deviates and grows into the grain interior along a plane close to the basal plane of the orange-coloured matrix grain. In addition, clear changes in crystal orientation, as compared to the bulk deuteride along the grain boundary, can be observed in those two regions.

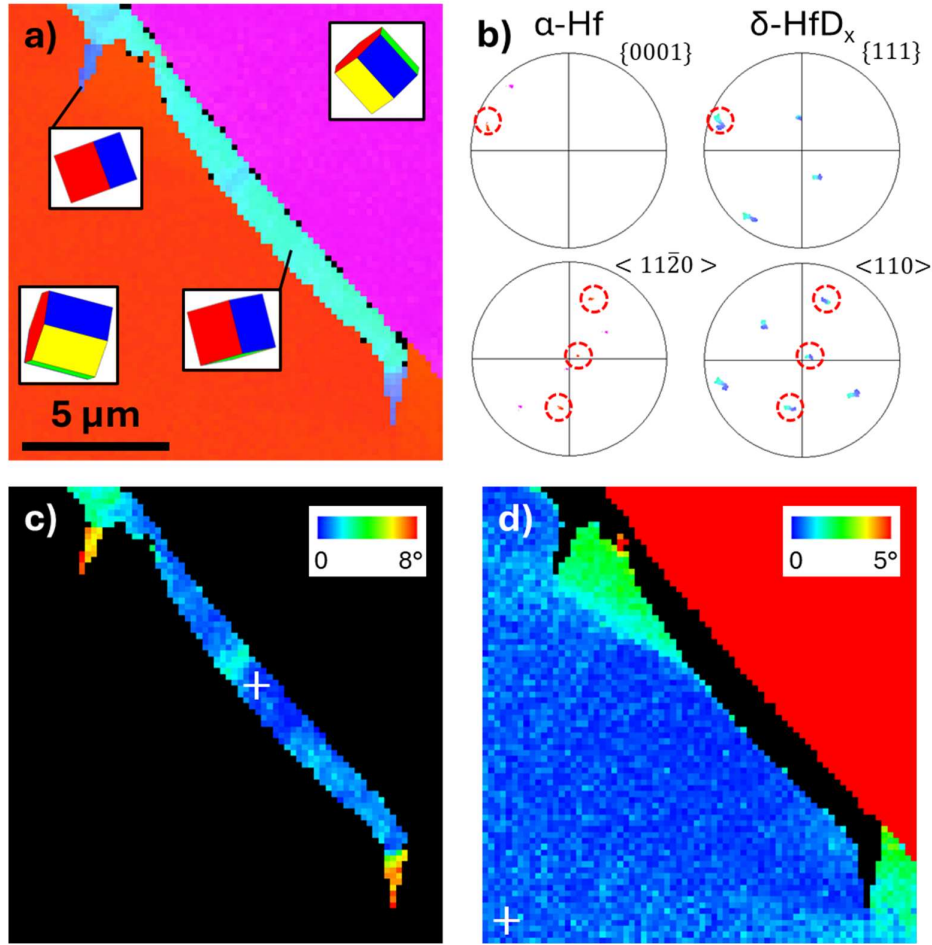

Figure 2 (a) Crystal orientation map, (b) pole figures, and (c,d) crystal misorientation to reference point (white crosses) maps of an intergranular deuteride packet.

Analysis of the orientation relationship (OR) between the deuteride and the α-Hf matrix is enabled by analysing the pole figures in Figure 2(b). The deuteride follows the $\{0001\}_\alpha \| \{111\}_\delta; <11\bar{2}0>_\alpha \| <110>_\delta$ OR with the orange-coloured matrix grain on the left side of the grain boundary, indicating that the deuteride packet was transformed from this matrix grain during the cooling process. Compared to other Group IV elements, this OR in Hf is the same as the predominant matrix-(δ-)hydride/deuteride OR in Zr [43,49–51] and one of the commonly observed matrix-(δ-)hydride/deuteride ORs in Ti [52,53].

Figure 2(c) and 2(d) show the misorientation maps relative to the reference point (the white cross in each image), highlighting the changes in crystal orientation within the deuteride and the surrounding matrix respectively. In Figure 2(c), it is observed that local misorientations within the deuteride develop as the deuteride deviates from the grain boundary. Notably, misorientations of up to 8° are evident in the protrusions. In Figure 2(d), local misorientations within the matrix are concentrated on the right side of the two protrusions in the deuteride.

A second example of intergranular deuteride is shown in Figure 3. Similar to the example in Figure 2, a deuteride packet is observed on one side of the grain boundary (the upper side in this case). Here, however, two distinct crystal orientations are visible within the deuteride (the green and red areas in Figure 3(a)). The pole figures in Figure 3(b) show that both deuteride orientations follow the $\{0001\}_\alpha \| \{111\}_\delta; <11\bar{2}0>_\alpha \| <110>_\delta$ OR with the pink-coloured matrix grain on the upper side of the grain boundary. In addition, the two deuteride orientations are $\{111\} <11\bar{2}>$ twins of each other, with the twin boundaries approximately parallel to the basal plane of the matrix.

Similar to the example in Figure 2, localised crystal misorientations (of up to 10°) are observed in the regions near the protrusions, as shown in Figure 3(c). Specifically, these localised misorientations are located near the centre of the map, slightly towards the top, and also on the bottom right. The deuteride-matrix interfaces at these protrusions are oriented close to the basal plane of the pink-coloured matrix. Additionally, Figure 3(c) reveals a 'deuteride-affected zone', where crystal misorientation is present in the matrix grain a few microns ahead of the deuteride growth front, as highlighted by the dashed line.

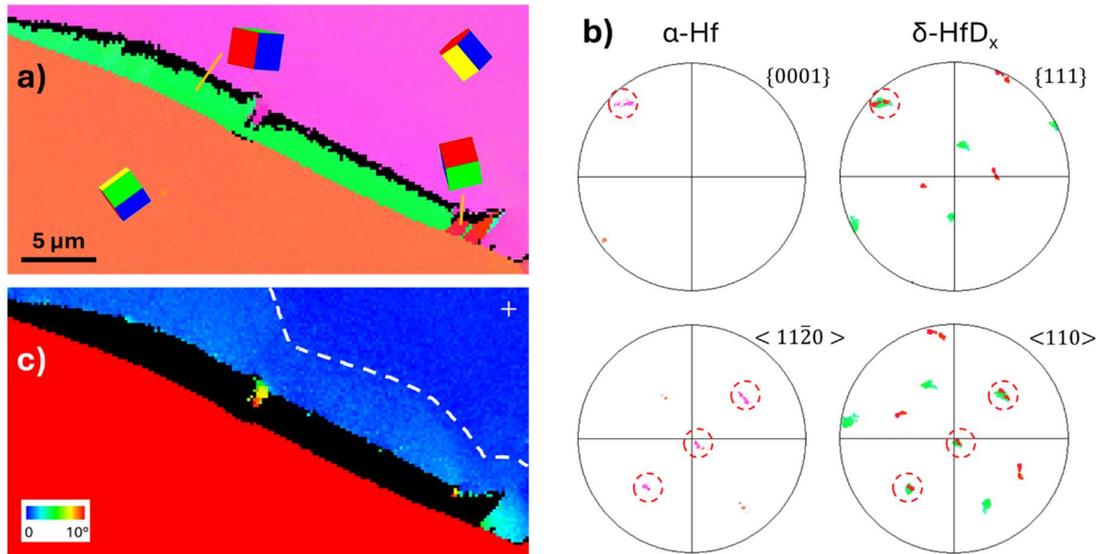

Figure 3 (a) Crystal orientation map, (b) pole figures, and (c) crystal misorientation to reference point (white cross) map of an intergranular deuteride packet. The deuteride packet contains twin structure.

Figure 4 illustrates an example where multiple crystal orientations are present in the deuteride on both sides of the grain boundary. The three deuteride orientations can be matched with the $\{0001\}_\alpha \| \{111\}_\delta; <11\bar{2}0>_\alpha \| <110>_\delta$ OR with their adjacent matrix grains, as can be seen in the pole figures. In Figure 4(b), the red- and blue-coloured deuterides, which both formed within the purple-coloured matrix grain and thus follow the OR with it, are $\{111\} <11\bar{2}>$ twins of each other; notably, the twin boundaries lie approximately parallel to the basal plane of the matrix, as was also observed in Figure 3. The green-coloured deuteride formed within the pink-coloured matrix grain on the right side of the map, as indicated by the OR between them.

Similar to Figures 2 and 3, for the deuteride packet in Figure 4, localised crystal misorientation is observed where the deuteride deviates from the grain boundary near the top right of the maps, as shown in Figure S3(b). In this case, the green-coloured deuteride in Figure 4(b) also grows along a plane close to the basal plane of the matrix after its deviation from the grain boundary.

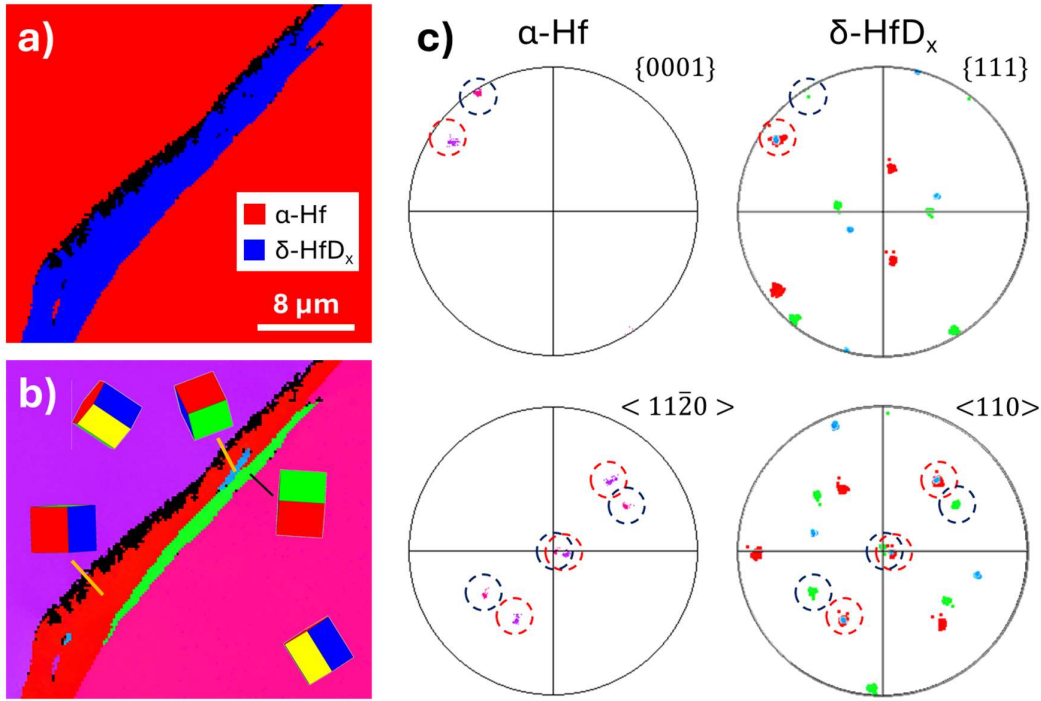

Figure 4 (a) Phase map, (b) crystal orientation map, and (c) pole figures of an intergranular deuteride packet. The deuteride packet contains deuterides formed in both grains and twin structure.

In total, we characterised such intergranular deuterides at 12 grain boundaries, and observed 21 different deuteride orientations (including deuterides on both sides of grain boundaries and deuteride twins). All these deuterides follow the same OR ($\{0001\}_\alpha \| \{111\}_\delta; <11\bar{2}0>_\alpha \| <110>_\delta$) with their adjacent matrix grains. Additionally, the pole figures show that the projections of the deuterides are generally more dispersive than those of the matrix. This suggests relatively higher internal lattice rotation gradients in the deuterides, likely due to a higher number of geometrically necessary dislocations within them. This is supported by the kernel average misorientation maps of these deuteride packets (Figure S2) where generally higher degrees of misorientations can be observed within the deuterides than in the matrix. Similar observations were made in the EBSD analysis of hydrides in Zr alloys [43].

We presented here a few representative deuteride packets to illustrate their microstructural and crystallographic characteristics. In the relatively large-grained material studied (Figure 1, average grain size is ~90 μm), the deuteride packets are widely spaced. Obtaining statistically meaningful data within a single EBSD scan (similar to our prior work on fine-grained Zr [43]) is practically challenging, as this would require both a sufficient number of deuterides and the fine step size needed for appropriate spatial resolution. It is also important to bear in mind that the motivation of studying hydrides in Zr is to prevent hydrogen embrittlement of fuel cladding through suppressing hydride formation. In such a case, statistical EBSD data analysis can provide knowledge about, for example, which type(s) of boundaries are more resistant to hydride nucleation, thereby informing potential alloy design to control boundary characteristics. Nonetheless, for Hf where the metal is proposed here to serve as a starting material for the fabrication of Hf-hydride composites, i.e., hydriding is desired rather than unwanted, statistical analysis becomes less relevant. Where there is a need for such analysis, future studies on fine-grained material, likely achievable by controlling heat treatment parameters, could help provide insight into the overall distribution of deuteride characteristics.

## 3.3. TEM/STEM characterisation of deuteride-matrix interfaces

An annular dark field (ADF) image of the growth front microstructure of an intergranular deuteride is shown in Figure 5(a1). In addition to the bulk intergranular deuteride on the right side of the image, needle-shaped intragranular deuterides embedded in the Hf matrix on the left can be observed. These intragranular deuterides may have formed during the 1 °C/min cooling process after hydrogen diffusion treatment, or they might have been induced by Xe plasma FIB milling (in which case they would more likely be hydrides). However, they will be referred to as 'intragranular deuterides' in this manuscript, given the uncertainty in their formation mechanism, which could be explored in future studies.

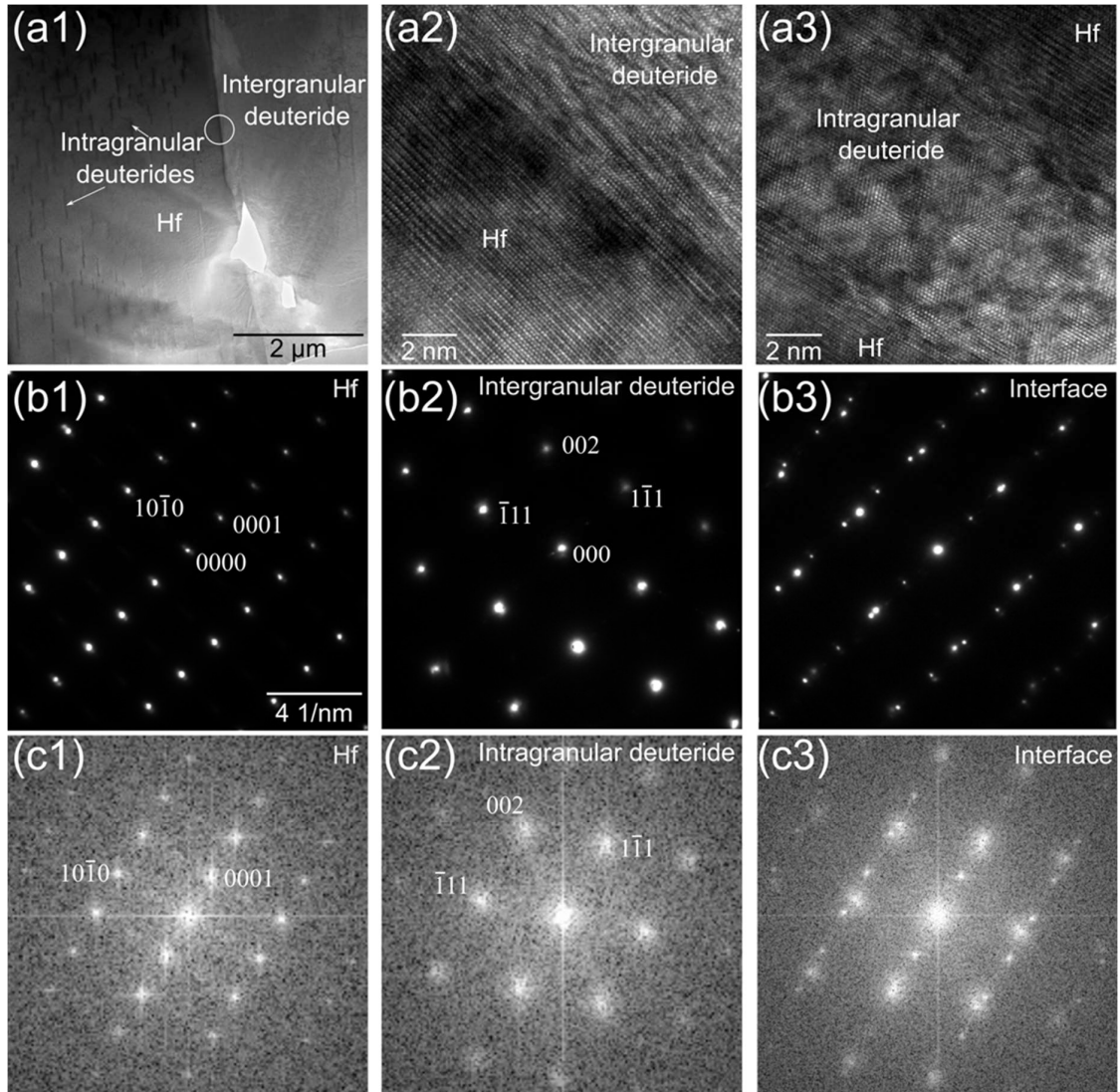

Figure 5 (a) TEM images of deuteride-matrix interfaces: (a1) ADF image showing the growth front of an intergranular deuteride, with thin intragranular deuterides visible in the Hf matrix. HR-TEM images of the interfaces between the matrix and (a2) the intergranular deuteride, and (a3) an intragranular deuteride. (b) SAED patterns of (b1) the matrix, (b2) the intergranular deuteride, and (b3) their interface, obtained from the circled region in (a1) when the sample was viewed in the orientation of (a2). (c) FFT of different regions in (a3): (c1) the matrix, (c2) the intragranular deuteride, and (c3) their interface.

High-resolution TEM (HR-TEM) images showing the interfaces between the matrix and both types of deuterides are presented in Figure 5(a2) (for intergranular deuteride) and (a3) (for intragranular deuteride). To determine the crystal structures and lattice parameters of these deuterides, selected area electron diffraction (SAED) was used for the intergranular deuteride (Figure 5(b2)), while for the thin intragranular deuteride, a fast Fourier transform (FFT) of a deuteride region in the HR-TEM image was used instead (Figure 5(c2)), as the size of the deuteride was smaller than the minimum available SAED aperture, making it impossible to isolate a diffraction pattern from the deuteride alone. These analyses reveal that both types of deuterides exhibit an FCC crystal structure with nearly identical lattice parameters (see Table S2), consistent with values for δ-deuteride derived from neutron diffraction [30].

The deuteride-matrix OR was determined by analysing SAED patterns for the intergranular deuteride (Figure 5(b1-b3)) and FFTs of different regions in the HR-TEM image for the intragranular deuteride (Figure 5(c1-c3)). Both types of deuterides are related to the matrix via the OR $\{0001\}_\alpha \| \{111\}_\delta$ (overlapping spots) and $<11\bar{2}0>_\alpha \| <110>_\delta$ (parallel zone axes), consistent with the EBSD results. Additionally, Figure 5(a2) and (a3) indicate that, for both types of deuterides, the interfaces are predominantly along the {0001} basal plane of the matrix and the {111} plane of the deuterides at the atomic scale.

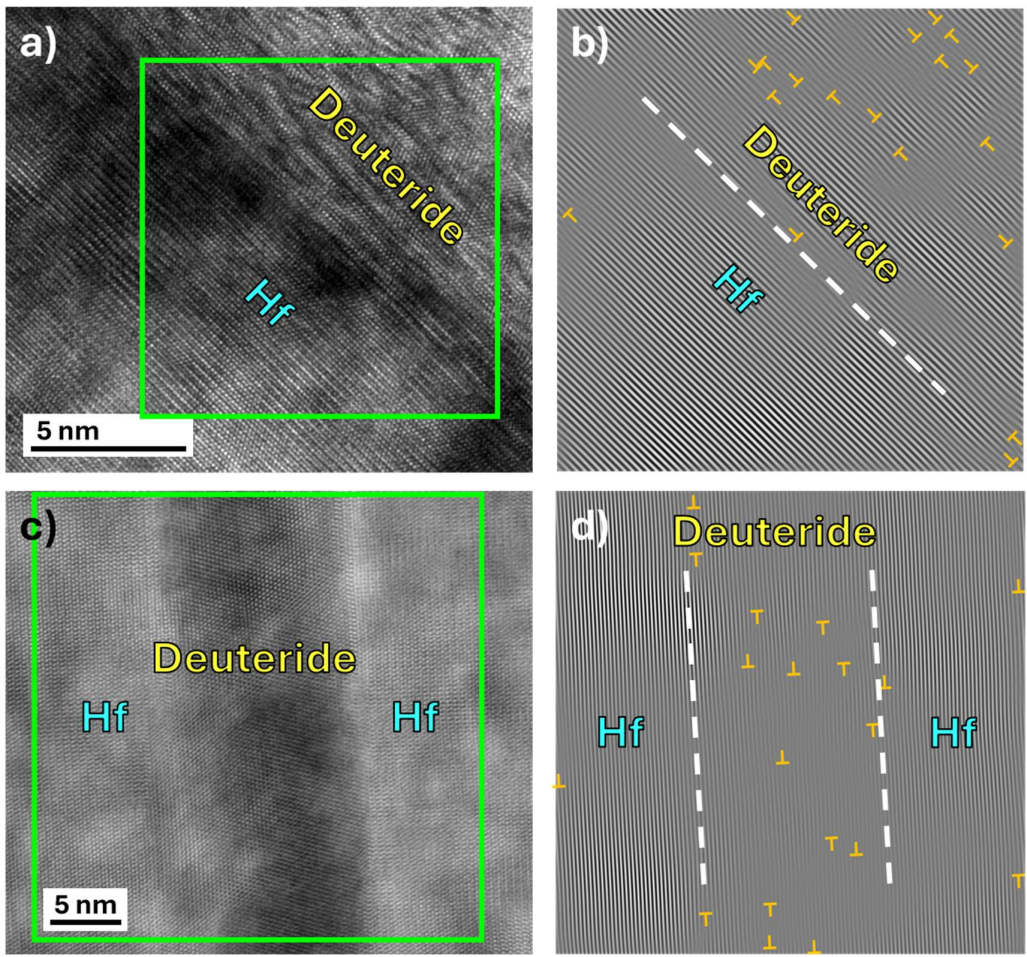

Figure 6 HR-TEM images of the interfaces between the matrix and (a) an intergranular deuteride and (c) an intragranular deuteride. (b) and (d) IFFT images, generated by isolating the overlapping $(0002)_\alpha/(111)_\delta$ pseudo diffraction spots in the FFTs of the regions within the green boxes in (a) and (c), respectively, revealing the real-space images of the atomic planes corresponding to these spots. Note that (a) is a TEM image, while (c) was acquired in STEM mode. Orange-coloured T-shaped marks in (b) and (d) highlight the positions of edge dislocations identified in the IFFT images. These IFFT images without the marks are provided in Figure S4.

To further investigate the lattice defects associated with deuteride formation, inverse fast Fourier transform (IFFT) analysis was performed on selected HR-TEM images, including one of an intergranular deuteride and one of an intragranular deuteride. This method isolates specific crystallographic information from the FFT, enabling the visualisation of dislocations and lattice distortions at the atomic scale. The results of this analysis are presented in Figure 6.

The IFFT images reveal a significantly higher number of dislocations (highlighted with orange T-shaped marks) in the deuteride packets compared to the surrounding matrix (unmarked versions of these images are provided in Figure S4). This is consistent with the EBSD results presented in Section 3.2, where the deuteride projections in the pole figures are generally more dispersive than the matrix projections, implying more dislocations in the deuteride.

Another notable feature of the dislocations in the deuterides, as observed in the IFFT images, is their relatively broad core structure, characterised by regions of lattice distortion surrounding the dislocations. One example of this is shown in detail in Figure S5. The dislocation illustrated has a Burgers vector $b = <\bar{1}12>$, and its width (defined as the distance between the two points where the disregistry transitions to zero, corresponding to the boundaries where the lattice returns to its perfect crystal structure) is approximately $24b$.

## 3.4. EELS analysis of deuteride-matrix interfaces

Figure 7 shows the results of the EELS analysis of the interfaces between the matrix and the intergranular deuteride (a1-a3) and the intragranular deuteride (b1-b3). The following observations can be made:

i. The plasmon peak energy for α-Hf is measured at 18.0 eV, while both intergranular and intragranular deuterides exhibit an identical plasmon peak energy of 20.0 eV. In Zr, we previously measured EELS plasmon peak energies for α-Zr, γ-hydride, and δ-hydride at 16.9 eV, 18.4 eV, and 19.0 eV respectively [54], consistent with other literature showing that γ- and δ-hydrides differ by ~0.6-0.9 eV [55,56]. Given these systematic and distinguishable energy shifts between distinct hydride phases in Zr and the well-documented similarities between Zr-H and Hf-H systems [30], the identical plasmon peak energy observed here for the intergranular and intragranular deuterides in Hf suggests that they belong to the same phase. This aligns with the phase identification from the crystal structure analysis in Section 3.3, which confirmed that both groups correspond to the δ phase.

ii. While the SI in (a2) shows a distinct deuteride-matrix interface for the intergranular deuteride (evidenced by a clear transition in plasmon peak energy), the interfaces between the matrix and the intragranular deuteride in (b2) exhibit a more gradual transition. This is likely due to the small size of the intragranular deuterides, which do not extend through the full thickness of the specimen, and so the EELS signal is due to overlap of signal from both the deuteride and the matrix. This is particularly noticeable towards the edges of the deuteride.

For α-Hf, the EELS spectra and plasmon peak energy (18.0 eV) obtained here using a 300 kV primary beam agree well with prior transmission-mode EELS results using a 20 kV primary beam, which also reported a plasmon peak energy of 18.0 eV [57]. In comparison, the bulk plasmon peak energy measured using reflective mode EELS with a 0.5 kV primary beam was moderately higher, at 19 eV, likely due to surface effects [58].

Theoretically predicted X-ray absorption near-edge structure (XANES) spectra of the Hf $L_{III}$ edge have been obtained for $HfH_{1.44}$ and $HfH_{1.99}$ using the real space multiple scattering code, FEFF [39], which is also capable of simulating EELS spectra [59,60]. Band structure calculations have also been carried out for $HfH_2$ using the WIEN2k code [61]. However, to the best of our knowledge we present the first experimental EELS spectra focusing on the plasmon region, which can be used to identify δ-deuterides in Hf, as has been shown for hydrides in Zr [55]. It is worth noting that, although the plasmon peak energies of the intergranular and intragranular structures are consistent, suggesting they are both δ phase, this may not confirm that the intragranular structures are deuterides (as opposed to hydrides, if they were FIB-induced). This uncertainty arises because it is unclear whether hydrides and deuterides would show a noticeable difference in plasmon peak energies due to the subtle mass difference between H and D. Thus, future studies comparing the EELS spectra for hydrides and deuterides could clarify whether this isotope effect of H/D is large enough to manifest in the plasmon peak energy in Hf and similar materials.

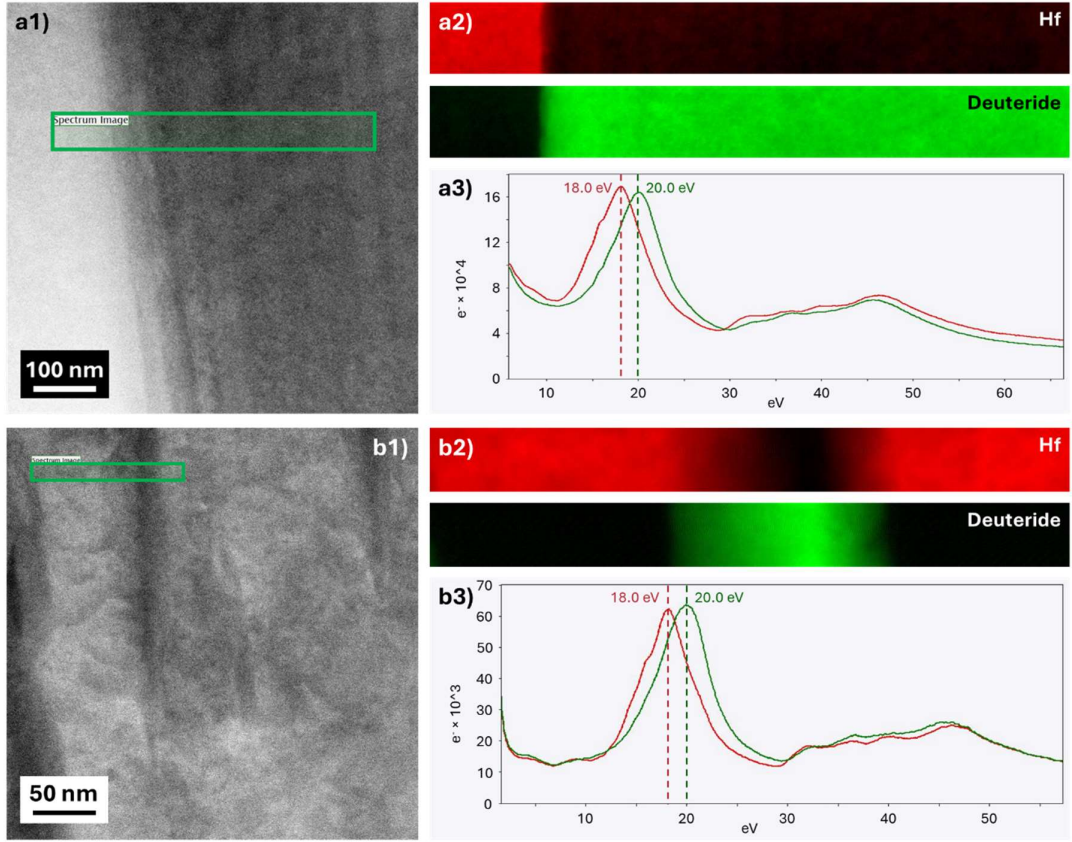

Figure 7 (a1) ADF image of the interface between the Hf matrix and the intergranular deuteride. (a2) Spectrum images (MLLS) for the two plasmon peaks in the Hf matrix and deuteride regions, both corresponding to the same area, highlighted with the green box in (a1). (a3) A 'standard' spectrum from each region. (b1-b3) show the same information as (a1-a3) respectively, for an intragranular deuteride.

### 3.5. APT analysis of deuteride growth front

Figure 8(a) presents the APT element maps obtained at the growth front of an intergranular deuteride. As a clarification, the deuterides analysed here using APT are the thick intergranular deuterides formed as a result of D charging, not the thin intragranular features discussed earlier, which may be FIB-induced hydrides. The Hf map highlights the contour of the APT needle, and an SEM image of the needle is provided in Figure S6.

The mass spectrum showing the H/D-related peaks is provided in Figure S7. The presence of peaks at 3 and 4 Da, corresponding to $HD^+$ and $D_2^+$ species respectively, further confirms the deuteride nature of the sample. Moreover, the relative intensities of the peaks in Figure S7 resemble those observed for Zr deuteride in our previous work [46]. As in that study, the peak at 1 Da, attributed solely to H, was excluded from the analysis of D content. The peak at 2 Da was assumed to arise entirely from $D^+$, with no contribution from $H_2^+$. While this assumption could lead to an overestimation of the D content if the 2 Da peak includes a partial $H_2^+$ contribution, precise D quantification is beyond the scope of this work and does not affect the main conclusions.

For O quantification, all O-containing species were included; over 90% of these were HfO ions, while monatomic O ions accounted for less than 1% of the total O signal, with a relatively higher concentration observed near the apex of the tip. It is worth noting that the dominance of HfO ions in the detected signal does not necessarily indicate that oxygen atoms were originally chemically bonded to hafnium in the material [62].

Composition profiles across the deuteride-matrix interface, traced along the black arrow in the D/O map in Figure 8(a), are shown in Figure 8(b). The distribution of O atoms reveals two notable features:

- A generally lower O concentration in the δ-deuteride compared to the α-Hf matrix.
- A distinct segregation of O atoms (reaching up to ~7%) at the deuteride-matrix interface.

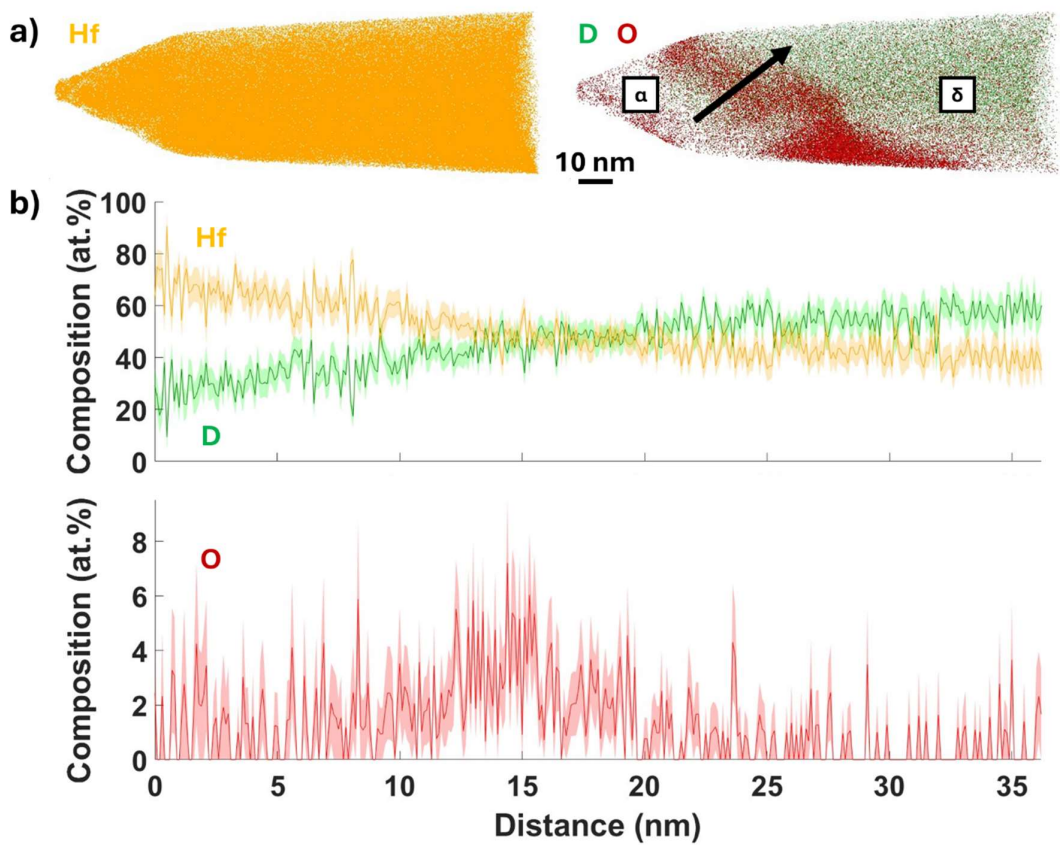

Figure 8 (a) APT element maps of the specimen shown in Figure S6, illustrating the distributions of Hf, D, and O. (b) Composition profiles of Hf, D, and O across the deuteride growth front along the black arrow on the D/O map in (a).

While the analysis described above provides a practical means of estimating D content, it is important to emphasise that APT data should not be used to derive the exact stoichiometry of the deuteride, even if the measured value appears reasonable in this case (the measured D content in the δ-deuteride is ~60%, corresponding to a stoichiometry of $HfD_{\sim 1.5}$, which aligns with values reported in the literature [29]). Accurate H/D quantification by APT is hindered primarily by the background H signal originating from the analysis chamber and the overlapping peaks of atomic D and molecular $H_2$ at 2 Da. Quantifying this overlap would require dedicated comparative studies on D-charged, H-charged, and uncharged Hf samples prepared using cryogenic Xe plasma FIB (assumed here to minimise FIB-induced hydrides, based on its demonstrated effectiveness in Zr [56,63]), along with careful data interpretation accounting for variations in electrostatic field strength within and between samples, as demonstrated in our previous work on Zr [44]. While such studies would be valuable for advancing H/D quantification in Hf using APT, they fall outside the scope of the present work, which aims to elucidate the structural and chemical features of the deuteride-matrix interfaces that underpin their formation and evolution. APT quantification of O is also subject to uncertainty, as the exact O content may be overestimated due to residual oxygen-containing gases in the analysis chamber [45]. However, this is unlikely to significantly affect the analysis of the deuteride-matrix interface largely situated within the sample interior, particularly the composition profile in Figure 8(b), which was taken along a trajectory well away from the surface, as gas-induced artefacts are most active at the surface (consistent with our observation of elevated monatomic O ions near the apex). These issues were discussed in more detail in our prior work on Zr [44–46], and further in-depth insights can be found in a recent publication [47].

## 4. Discussion

### 4.1. Microstructural characteristics at deuteride-matrix interfaces

Characterisation of the deuterides using EBSD and TEM/STEM has revealed the following characteristics:

i. The presence of dislocations (and hence crystal orientation gradients at a larger scale), and the number of dislocations is evidently higher in the deuteride than in the matrix.
ii. The frequent formation of $\{111\} <11\bar{2}>$ (or Σ3) twin structures in the deuterides, with twin boundaries approximately parallel to the basal plane of the matrix.
iii. The local growth of deuterides along a plane (the 'habit plane') close to the basal plane of the HCP α-Hf matrix.

These characteristics are highly similar to those observed in the Zr-hydride system. In Zr, a higher dislocation density in the hydride compared to the matrix has been observed using various techniques, including synchrotron X-ray diffraction [64], neutron diffraction [50], and electron channelling contrast imaging [46], and it is generally accepted that these dislocations are produced upon hydride formation to accommodate the lattice misfit [43,50,64]. The formation of Σ3 twins in hydrides in Zircaloy-4 is related to the atomic-scale nature of the phase transformation, and a detailed discussion of the potential mechanism is provided in our prior work [43]. The habit plane of hydrides in Zr is usually close to the basal plane of the matrix where atomic matching occurs due to the OR. However, continued growth strictly along the basal plane is energetically unfavourable because of accumulating misfit strain. As a result, the habit plane is often slightly tilted away from the exact basal plane to reduce the overall transformation strain energy.

The lattice misfit along specific crystallographic directions can be estimated using the deuteride-matrix OR and the lattice parameters of the two phases (Table S2). For example, along the $<10\bar{1}0>_\alpha \parallel <112>_\delta$ direction, the lattice parameters are 5.51 Å for α-Hf and 5.77 Å for δ-deuteride, corresponding to a misfit of 4.7%. This suggests that, on average, one dislocation is required every ~21 atoms to accommodate the mismatch. This estimation is consistent with the broad dislocation core structure observed in HR-TEM (Figure S5), where the core width of a representative dislocation was measured to be approximately $24b$. Such an extended core likely reflects the gradual relief of lattice misfit over a long distance, facilitated by the displacive transformation mechanism that proceeds atom by atom. Therefore, in hydride-forming HCP metals, the width and density of misfit-induced dislocations are likely fundamentally governed by the local lattice strain resulting from the precise crystallographic mismatch between the two phases.

The only deuteride phase observed in this work is the FCC δ-deuteride. Another stable hydride (deuteride) phase in the Hf-H(D) system, as introduced in Section 1, is the FCT ε phase (HfH(D)$_{\leq 2}$) which can be transformed from the δ phase as hydrogen (deuterium) concentration increases [29]. In both phases, the H(D) atoms occupy the tetrahedral interstitial sites of the FCC(FCT) Hf lattice. First-principles calculations of the electronic structure and energetics of HfH$_2$ (i.e., the ε phase) showed that, for this stoichiometry where all the tetrahedral sites are fully filled, cubic structure is unstable and the ground state of HfH$_2$ corresponds to a tetragonal structure with *c/a* < 1 [61], in agreement with experimental observations. The tetragonal distortion of the ε phase, and the resulting changes in its electronic structure, is expected to produce EELS features that differ from those of the δ phase, as also implied by photoemission results on ZrH$_x$, which showed changes in Fermi-level emission and binding energy associated with the similar FCC-FCT distortion [65]. Experimental validation of this would also facilitate phase identification in the Hf-H system.

## 4.2. Mesoscale deuteride (hydride) growth mechanism

In Zr and Ti, hydride formation is generally accepted to occur *via* the following mechanisms:
  a. At the microscale and especially for intragranular hydrides, repeated auto-catalytic nucleation [49,66] results in near-basal habit planes [67], as discussed in Section 4.1.
  b. At a larger (typically sub-mm) scale, in fine-grained metals, hydrides grow across grain boundaries also *via* an auto-catalytic mechanism [43] to form interconnected stringers.

There is, however, still limited understanding of hydrides' growth patterns in between the above two length scales, i.e., at the mesoscale, shortly after nucleation at preferential sites such as grain boundaries, how hydrides grow into the complicated morphologies observed in Figures 2-4 and in Zr [43] and Ti [68], as likely driven by the competition between the chemical potential well at the nucleation site and that around the hydride embryo.

Our results show that despite the formation of local protrusions, most intergranular deuterides do not penetrate deeply into the grain interior but instead decorate grain boundaries at a larger scale (see Figure 1(b)). This morphology can be attributed to dislocations initially formed around the side walls of the protrusions (Figures 2, 3, S2): the elastic strain fields associated with these dislocations attract solute deuterium (hydrogen) atoms, promoting deuteride precipitation once the local solubility limit is exceeded [43]. Thus, although growth along the habit plane may be energetically favourable initially, the associated dislocation formation progressively creates a driving force that impedes further protrusion. Instead, deuteride formation is redirected into adjacent gaps near the protrusions.

Therefore, the thickening process of intergranular deuterides likely proceeds through a 'protrusion-sweeping' mechanism depicted in Figure 9. Each deuteride structure characterised by EBSD may represent a distinct stage in this process: for example, the deuteride in Figure 2 likely corresponds to an early stage with freshly formed protrusions, while the one in Figure 3 depicts a later stage in which sweeping has nearly completed in the central region of the map. Confirming the proposed mechanism unambiguously would ideally require time-resolved or *in situ* experimental validation. For example, *in situ* high-temperature EBSD, which we recently used to study phase transformations in solders [69], could be extended to Hf-H and Zr-H systems in future work. In addition, studying the *in situ* formation (and potentially the dissolution or desorption) of hydrides with *in situ* transmission electron microscopy and/or X-ray diffraction could also generate results in support of the *in situ* neutron diffraction experiments of Dottor et al. [23]. Together these could strengthen and give insight into the underlying mechanisms of hydride behaviour.

Although established models of hydride growth in Hf are scarce, our proposed mechanism is in remarkable agreement with that reported recently by Liu *et al.* [70], who used a density functional theory-informed crystal plasticity finite element model to simulate hydride precipitation in Zr. In their simulation, intergranular hydrides grow *via* a two-step process: planar growth parallel to the basal planes, driven by the high compressive stress within these planes that constrains the hydrides to grow in-plane initially, followed by stacking growth perpendicular to the basal planes, driven by α-δ interactions and uneven hydrogen distribution around hydride platelets. This consistency suggests that hydride formation in Hf is likely very similar to that in Zr in crystallographic and microstructural aspects, although differences exist in interfacial chemistry (see Section 4.3).

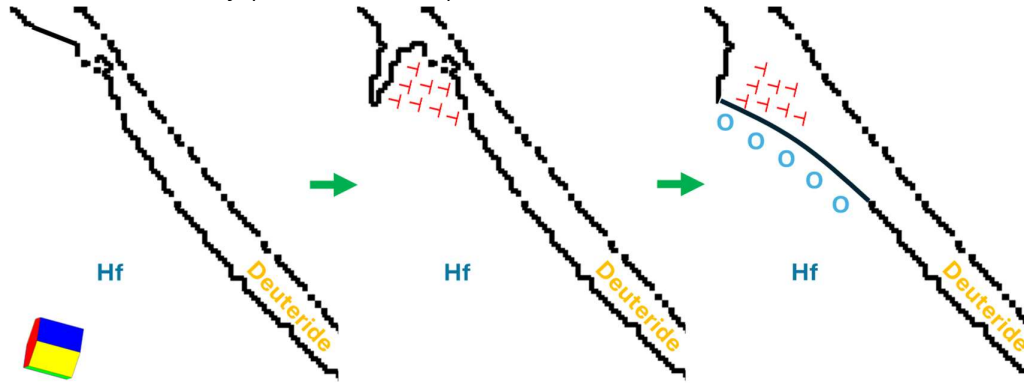

Figure 9 Schematic illustration of the thickening process of an intergranular deuteride, formed in a grain where the basal plane is oriented at an angle to the grain boundary.

Furthermore, as deuterides grow, they necessarily incorporate the misfit-induced dislocations originally present in the matrix into the advancing deuteride phase (Figure 9), which likely contributes to the higher dislocation density observed within the deuterides. One possible alternative explanation is that the deuteride phase is intrinsically softer than the matrix, and therefore more prone to plastic deformation under misfit stress. However, this interpretation is not supported by experimental data. In the Zr-hydride system, for instance, hydrides have been shown to be stronger than the surrounding matrix [71–73]. Similarly, in Hf, using the same material investigated in this work, we conducted *in situ* compression tests on deuteride-matrix composite micropillars. These tests consistently revealed slip bands in the matrix that were arrested at the deuteride-matrix interfaces (Figure S8), indicating that the matrix is comparatively softer and more readily accommodates plastic deformation than the deuteride.

Unlike dislocations, which can be absorbed by the deuteride/hydride during growth, oxygen atoms are rejected from the deuteride and accumulate ahead of the advancing interface, as revealed by APT data. While deuteride growth can still proceed in the presence of interfacial oxygen, its continued accumulation may pose increasing resistance, potentially acting as a solute drag that slows interface migration and eventually limits further growth. In addition, oxygen may induce local lattice distortions at the interface, further influencing transformation kinetics. The suppression of growth may lead to a finer distribution of deuterides/hydrides in the microstructure. This could influence key material properties, such as toughness, neutron moderation, and thermal conductivity. Hence, oxygen content and distribution may have important implications for the microstructure and, consequently, the in-service performance of these materials. As such, careful control of oxygen content during manufacturing may be necessary. Further discussion of oxygen's effects on structural integrity is provided in Section 4.3.

It is worth noting that the results presented in this work do not imply that oxygen necessarily plays a role in potential hydride dissolution or hydrogen desorption processes, which may occur when the material is exposed to elevated temperatures during reactor service. Clarifying the microstructural and atomic-scale characteristics of a dissolving or desorbing interface, using a material such as that studied in Ref. [27] and applying the same techniques employed in the present work, would help advance our understanding of the in-service performance of these materials further.

## 4.3. Implications on structural integrity

As discussed in Section 4.1, deuterides tend to grow along habit planes that deviate slightly from the basal plane of the HCP matrix due to lattice misfit, which leads to significant stress concentrations near the protrusions. If this habit plane deviation is driven by misfit, its angle is expected to decrease with decreasing mismatch. Alloying may offer a route to tune the misfit and influence the coherency and orientation of the interface. For instance, reducing the lattice mismatch could promote a habit plane that more closely aligns with the basal plane of the matrix, potentially enhancing interfacial coherency and mitigating stress concentrations. Such an approach may be beneficial for improving the mechanical integrity of Hf deuteride/hydride systems, and may also be extended to other hydride-forming HCP alloys such as Zr, where hydride embrittlement in fuel cladding remains a critical concern.

Compared to pure Hf hydride, Hf-hydride composites are expected to offer improved toughness and ductility due to the mechanical support of the metallic matrix. However, as in analogous Zr and Ti systems [73–75], failure can still occur, typically initiating at hydride-matrix interfaces under stress. A key difference in Hf is the observation of oxygen segregation at these interfaces - an effect not reported in previous APT studies of Ti [76] or Zr [46], aside from some enrichment at defects in Zircaloy-4 [45]. This segregation may compound the interfacial strain localisation, potentially promoting failure *via* mechanisms analogous to oxygen-induced intergranular fracture observed in metals such as Fe [77], Cu [78], and Mo [79]. Thus, beyond influencing hydride microstructure (Section 4.2), oxygen may also directly affect mechanical integrity through its interfacial presence. While these effects remain to be confirmed by micromechanical testing, our findings suggest that oxygen control could be key to ensuring the structural reliability of these materials under reactor conditions.

## 4.4. Deuterides/hydrides or 'FCC Hf/Zr/Ti'?

Our APT results confirm the deuteride nature of the precipitates. However, in previous TEM studies of Hf, Zr, and Ti alloys – where hydrogen atoms are invisible even when present – similar FCC-structured features have been interpreted either as hydrogen-free metallic phases or as hydrides [80,81], leading to an ongoing debate over their true nature. Distinguishing between these possibilities is crucial: in failure analysis, misidentifying a hydride as an FCC metal (or vice versa) could lead to incorrect conclusions about the underlying mechanisms; in alloy design, it could result in critical misjudgements – for example, assuming a desirable ductile FCC metallic phase has formed when it is, in fact, a brittle hydride that compromises performance.

In practice, however, resolving this distinction is challenging, primarily due to

i. The high cost and, consequently, limited accessibility of equipment capable of directly resolving or detecting hydrogen atoms, such as advanced TEM [82] and APT, and
ii. The general lack of knowledge and standardised protocols for preparing appropriate 'FCC Hf/Zr/Ti' samples.

While direct detection of hydrogen remains technically demanding, EELS offers a relatively accessible and effective alternative for distinguishing between deuterides/hydrides and the matrix. In Section 3.4, we demonstrated that the plasmon peak energy differs across an interface between the δ-deuteride and the α-Hf, suggesting that EELS can serve as an indicator of hydrogen presence in these systems. This observation is consistent with the Drude model, which relates plasmon peak energy to the square root of the valence electron density [83]. Importantly, the valence electron density, and therefore the plasmon peak energy, of a hydrogen-free FCC phase is expected to be moderately lower than that of the HCP phase [80]. In contrast, hydrides typically exhibit higher valence electron densities due to the contribution of hydrogen atoms, resulting in noticeably higher plasmon peak energies [80]. Therefore, comparing plasmon peak energies can, in principle, help distinguish deuterides/hydrides from the 'FCC Hf/Zr/Ti' phases.

Although EELS data for the hydrogen-free FCC Hf/Zr/Ti phases are, to the best of our knowledge, not yet available, reliable plasmon energy values for their deuteride/hydride counterparts have been reported in the literature (e.g., Ti [80,84–87], Zr [54–56,88], and Hf as reported in Section 3.4). We therefore recommend that future studies reporting hydrogen-free FCC phases include EELS spectra and compare the measured plasmon peak energies against those of known deuteride/hydride phases. If the spectra differed significantly from the deuterides/hydrides and were instead more consistent with those of the corresponding HCP metals, this would provide strong support for the interpretation that the observed FCC phases are indeed hydrogen-free.

# 5. Conclusions

We applied SEM-EBSD, TEM/STEM, EELS, and APT to characterise the microstructural and atomic-scale features of deuterides and their interfaces with the matrix in a Hf alloy, aiming to improve understanding of these features in materials designed primarily for nuclear applications. Key findings are as follows:

1. The deuterides follow a $\{0001\}_\alpha \| \{111\}_\delta; <11\bar{2}0>_\alpha \| <110>_\delta$ OR with the matrix. Their habit plane is close to the basal plane of the matrix, and they develop internal Σ3 twins during growth. These features closely resemble those in the Zr-hydride system.
2. Dislocations accommodate the lattice misfit between the deuteride and the matrix. The dislocation density and core structure, and the precise deuteride habit plane likely depend on the magnitude of the misfit.
3. We propose a 'protrusion-sweeping' mechanism to describe how intergranular deuterides thicken during growth. This mechanism may also explain the higher dislocation density observed within the deuterides.
4. Deuterides reject oxygen atoms, which then accumulate ahead of the growth front. This accumulation may affect the growth kinetics and microstructure of the deuterides. It may also compromise the structural integrity of the material.
5. To help resolve the long-standing controversy surrounding the so-called 'FCC Hf/Zr/Ti' phases, we recommend using EELS, particularly plasmon peak energy analysis, to distinguish hydrogen-containing phases from hydrogen-free ones.

To envisage future research directions, the following efforts may further clarify the material's suitability for nuclear applications:

a. Fabrication of materials more representative of actual reactor components, i.e., with higher hydride/deuteride fractions than that analysed here, *via* the industrial routes described in Section 1;
b. Conducting *in situ* experiments under conditions relevant to service environments to elucidate microstructural and chemical evolution, such as hydride/hydrogen loss which is a primary concern for hydride-based materials [39].
c. Establishing knowledge about the material's mechanical performance, including fundamental properties such as fracture toughness (currently elusive even for pure Hf hydride) and the effects of neutron irradiation on structural integrity, to assess the material's ability to withstand harsh operational environments.

# Acknowledgements

SW acknowledges Imperial College London for funding his Imperial College Research Fellowship. The TFS Quanta SEM used was supported by the Shell AIMS UTC and is housed in the Harvey Flower EM suite at Imperial College London. The Hydra dual beam plasma FIB-SEM, the Spectra 300 STEM, and the LEAP 5000, are part of the cryo-EPS facility at Imperial College London funded by Engineering and Physical Sciences Research Council (EP/V007661/1). The Matlab code used to plot unit cells for representation of crystal orientations was written by Dr Vivian Tong at National Physical Laboratory UK.wrong tagin the Harvey Flower EM suite at Imperial College London. The Hydra dual beam plasma FIB-SEM, the Spectra 300 STEM, and the LEAP 5000, are part of the cryo-EPS facility at Imperial College London funded by Engineering and Physical Sciences Research Council (EP/V007661/1). The Matlab code used to plot unit cells for representation of crystal orientations was written by Dr Vivian Tong at National Physical Laboratory UK.